\documentstyle[titlepage,12pt]{article}
\begin{document}
\renewcommand{\theequation}{\thesection.\arabic{equation}}
\title{Mass Spectrum of Strings in Anti de Sitter
Spacetime}
\author{A.L. Larsen and N. S\'{a}nchez\\
\\
\\
Observatoire de Paris,
DEMIRM. Laboratoire Associ\'{e} au CNRS \\
\hspace*{-9mm}UA 336, Observatoire de Paris et
\'{E}cole Normale Sup\'{e}rieure. \\
\hspace*{-32mm}61, Avenue de l'Observatoire, 75014 Paris, France.}
\maketitle
\begin{abstract}
We perform string quantization in anti de Sitter (AdS) spacetime. The string
motion is stable, oscillatory in time with real frequencies $\omega_n=
\sqrt{n^2+m^2\alpha'^2H^2}$ and the string size and energy are bounded. The
string fluctuations around the center of mass are well behaved. We find the
mass formula which is also well behaved in all regimes. There is an
{\it infinite}
number of states with arbitrarily high mass in AdS (in de Sitter (dS) there is
a {\it finite} number of states only). The critical dimension at which the
graviton appears is $D=25,$ as in de Sitter space.
A cosmological constant $\Lambda\neq 0$ (whatever its sign) introduces a
{\it fine structure} effect (splitting of levels) in the mass spectrum at all
states beyond the graviton. The high mass spectrum changes drastically
with respect to flat Minkowski spacetime. For $\Lambda<0,$ we find
$<m^2>\sim \mid\Lambda\mid N^2,$ {\it independent}
of $\alpha',$ and the level spacing
{\it grows} with the eigenvalue of the number operator, $N.$ The
density of states $\rho(m)$
grows like $\mbox{Exp}[(m/\sqrt{\mid\Lambda\mid}\;)^{1/2}]$ (instead of
$\rho(m)\sim\mbox{Exp}[m\sqrt{\alpha'}]$ as in Minkowski space), thus
{\it discarding} the existence of a critical string temperature.

For the sake of completeness, we also study the quantum strings in the
black string background, where strings behave, in many respects, as in
the ordinary black hole backgrounds. The mass spectrum is equal to the
mass spectrum in flat Minkowski space.
\end{abstract}
\section{Introduction and Results}
String dynamics in curved spacetime and its associated physical phenomena
started to be systematically studied in Refs.\cite{san1,san2}. Since then, this
subject has recieved systematic and increasing attention. String propagation in
curved spacetime reveals new insights and new physical phenomena with respect
to
string propagation in flat spacetime (and with respect to quantum fields in
curved spacetime). The results of this programme are relevant both for
fundamental (quantum) strings and for cosmic strings, which behave essentially
in a classical way. Approximative [1-5] and exact [5-10]
string methods have been developed.
Classical and quantum string dynamics have been investigated in
black hole backgrounds [5, 11-14], cosmological spacetimes \cite{ven}, cosmic
string spacetimes \cite{san9},
gravitational wave backgrounds \cite{veg1}, supergravity backgrounds
\cite{veg2} (which are necessary
for non-trivial propagation of fermionic strings) and near spacetime
singularities \cite{veg3}. Physical
phenomena like the Hawking-Unruh effect in string
theory \cite{san2,san10}, horizon string stretching \cite{san2,san10},
particle transmutation \cite{san7,veg4}, string scattering \cite{san7,san9},
mass spectrum and critical dimension \cite{san1,san7,san9},
string instabilities with non-oscillatory
motion in time [1, 5-9, 15], and multi-string solutions [7-9]
have been found. The physically rich
dynamics of strings in $D>2$ curved spacetimes is lost in $D=2,$ in which most
of the interesting stringy effects referred above disappear. String propagation
in the $D=2$ stringy black hole \cite{wit} reduces to the quantization of a
massless scalar particle (the dilaton) with a new peculiar effect: the
quantum
renormalization of the speed of light \cite{veg5}, which appears restricted to
strings in two dimensional curved spacetimes.

Recently \cite{all1}, the classical string dynamics was studied in
the $2+1$ black hole anti de Sitter
spacetime as well as in its dual, the black string background, within the
string perturbation series approach of de Vega and Sanchez \cite{san1}.
The first and second order string fluctuations around the center of mass were
obtained.
Comparison was made to the ordinary ($D\geq 4$) black hole backgrounds (with
cosmological constant). The circular string motion was exactly and completely
solved (in terms of either elementary or elliptic functions) in all these
backgrounds.

In the present paper we go further in the investigation of the physical
properties of strings in these backgrounds, by performing string quantization.
Since the ($2+1$) black hole spacetime is asymptotically anti de Sitter
(AdS), the results of Ref.\cite{all1}
can be extended to classical strings
in $D$-dimensional anti de Sitter spacetime as well. For the sake of
completeness, we also investigate string quantization in the black string
background, for which, in many respects, strings behave
as in the ordinary black hole backgrounds. In AdS
spacetime, the string motion is oscillatory in time and is {\it stable}; all
fluctuations around the string center of mass are well behaved and bounded.
Local
gravity of AdS spacetime
is always negative and string instabilities do not develop.
The string perturbation series approach, considering fluctuations around the
center of mass, is particularly appropriate in AdS spacetime,
the natural
dimensionless expansion parameter being $\lambda=\alpha'/l^2>0,$ where
$\alpha'$
is the string tension and the `Hubble constant' $H=1/l.$ The
negative cosmological
constant of AdS spacetime is related to the `Hubble constant' $H$ by
$\Lambda=-(D-1)(D-2)H^2/2.$

All the spatial ($\mu=1,...,D-1$) modes in $D$-dimensional AdS
oscillate with frequency
$\omega_n=\sqrt{n^2+m^2\alpha'^2H^2}=\sqrt{n^2+m^2\alpha'\lambda},$ which are
real for all $n$ ($m$ being the string mass). In this paper, we
perform a canonical
quantization procedure. From the conformal generators $L_n,\;\tilde{L}_n$ and
the constraints $L_0=\tilde{L}_0=0,$ imposed at the quantum level, we obtain
the mass formula:
\begin{equation}
m^2\alpha'=(D-1)\sum_{n>0}\Omega_n(\lambda)
+\sum_{n>0}\Omega_n(\lambda)\sum_{R=1}^{D-1}[(\alpha^R_n)^{\dag}
\alpha^R_n+(\tilde{\alpha}^R_n)^{\dag}\tilde{\alpha}^R_n],
\end{equation}
where:
\begin{equation}
\Omega_n(\lambda)=\frac{2n^2+m^2\alpha'\lambda}{\sqrt{n^2+m^2\alpha'\lambda}},
\end{equation}
and we have applied symmetric ordering of the operators. The operators
$\alpha^R_n,\;\tilde{\alpha}^R_n$ satisfy:
\begin{equation}
[\alpha^R_n,\;(\alpha^R_n)^{\dag}]=[\tilde{\alpha}^R_n,\;(\tilde{\alpha}^R_n)^
{\dag}]=1,\;\;\;\;\;\;\mbox{for all}\;\;n\geq 0,
\end{equation}
and we have eliminated the zero-modes. To the
first term in the mass formula (the zero point energy) we apply zeta-function
regularization, see Eqs.(3.14), (3.15). For $\lambda<<1,$ which is clearly
fulfilled in most interesting cases, we find the lower mass states
$m^2\alpha'\lambda<<1$ and the quantum mass spectrum. Physical states are
characterized by the eigenvalue of the number operator:
\begin{equation}
N=\frac{1}{2}\sum_{n>0} n\sum_{R=1}^{D-1}[(\alpha^R_n)^{\dag}
\alpha^R_n+(\tilde{\alpha}^R_n)^{\dag}\tilde{\alpha}^R_n],
\end{equation}
and the ground state is defined by:
\begin{equation}
\alpha^R_n\mid 0>=\tilde{\alpha}^R_n\mid 0>=0,\;\;\;\;\;\;
\mbox{for all}\;\;n>0.
\end{equation}
We find that $m^2\alpha'=0$ is an {\it exact} solution
of the mass formula in $D=25$ and that there is a graviton at $D=25,$ which
indicates, as in de Sitter space \cite{san1}, that the critical dimension
of AdS is 25 (although it should be stressed that the question whether de
Sitter space is a solution to the full $\beta$-function equations remains
open).
The ground state is a tachyon, its mass is given by Eq.(3.18).
Remarkably enough, for $N\geq 2$ we find that a generic feature of all excited
states beyond the graviton, is the presence of a {\it fine structure} effect:
for a given eigenvalue $N\geq 2,$ the corresponding states have different
masses. For the lower mass states the expectation value of the mass operator in
the corresponding
states (generically labelled $\mid {j}>$) turns out to have the
form, Eq.(3.32):
\begin{equation}
<{j}\mid m^2\alpha' \mid {j}>_{\mbox{AdS}}=a_{{j}}
+b_{{j}}\lambda^2+c_{{j}}\lambda^3+{\cal O}(\lambda^4).
\end{equation}
The collective index "j" generically labels the state $\mid {j}>$ and the
coefficients $a_{{j}},\;b_{{j}},\;c_{{j}}$ are all well computed numbers,
different for each state. The corrections to the mass in Minkowski spacetime
appear to order $\lambda^2.$ Therefore, the leading Regge trajectory for
the lower mass states is:
\begin{equation}
J=2+\frac{1}{2}m^2\alpha'+{\cal O}(\lambda^2).
\end{equation}
In Minkowski spacetime the mass
and number operator of the string are related by
$m^2\alpha'=-4+4N.$ In AdS (as well as in de Sitter (dS)), there is no such
simple relation between the mass and the number operators; the splitting of
levels increases considerably for very large $N.$ The {\it fine structure}
effect we find here is also present in de Sitter space.
Up to order $\lambda^2,$ the lower mass states in dS
and in AdS are the same, the differences appear to the order $\lambda^3.$ The
lower mass states in de Sitter spacetime are given by Eq.(1.6) but with the
$\lambda^3$-term getting an opposite sign ($-c_{{j}}\lambda^3$).

For the very high mass spectrum, we find more drastic effects. States with
very large eigenvalue $N,$ namely $N>>1/\lambda,$ have masses (Eq.(3.33)):
\begin{equation}
<j\mid m^2\alpha' \mid j>_{\mbox{AdS}}\approx d_{{j}}\lambda N^2
\end{equation}
and angular momentum:
\begin{equation}
J^2\approx\frac{1}{\lambda}m^2\alpha',
\end{equation}
where $d_{{j}}$ are well computed numbers different for each state.
Since $\lambda=\alpha'/l^2,$ we see from Eq.(1.8) that the masses of the high
mass states are {\it independent} of $\alpha'.$
In Minkowski spacetime, very large $N$ states all have the same mass
$m^2\alpha'\approx 4N,$ but here in AdS the masses of the
high mass states with the same
eigenvalue $N$ are {\it different} by factors $d_{{j}}.$
In addition, because of the fine structure effect, states with different $N$
can get mixed up. For high mass states, the level spacing {\it grows} with $N$
(instead of being constant as in Minkowski spacetime). As a consequence, the
density of states $\rho(m)$ as a function of mass grows like
$\mbox{Exp}[(m/\sqrt{\mid\Lambda\mid}\;)^{1/2}]$ (instead of
$\mbox{Exp}[m\sqrt{\alpha'}]$ as in Minkowski spacetime), and
{\it independently}
of $\alpha'.$ The partition function for a gas of strings at a temperature
$\beta^{-1}$ in AdS spacetime
is well defined for all finite temperatures $\beta^{-1},$
discarding the existence of the
Hagedorn temperature and the possibility of a phase
transition (as occurs in Minkowski spacetime and in other curved spacetimes).

In the black string background, we calculate explicitly the first and second
order string fluctuations around the center of mass. We then determine the
world-sheet energy-momentum tensor and we derive the mass formula in the
asymptotic region. The mass spectrum is equal to the mass spectrum in flat
Minkowski spacetime. Therefore, for a gas of strings at temperature
$\beta^{-1}$ in the asymptotic region of the black string background,
the partition function can only be defined for $\beta>\sqrt{\alpha'},$ i.e.
there is a Hagedorn temperature, Eqs.(4.51), (4.52).

This paper is organized as follows: In Section 2 we review the clasical string
propagation in the $2+1$ black hole anti de Sitter spacetime \cite{all1} and we
generalize the results to ordinary $D$-dimensional anti de Sitter spacetime.
In Section 3 we perform a canonical quantization of strings in
$D$-dimensional anti de Sitter spacetime and we analyze the string spectrum. In
Section 4 we consider classical and quantum string propagation in the
black string background. A summary of our results and conclusions is presented
in Section 5 and in Table 1.
\section{Classical String Propagation}
\setcounter{equation}{0}
The classical propagation of strings in the $2+1$ dimensional black hole anti
de Sitter spacetime was considered in detail in Ref.\cite{all1}
using both approximative
and exact methods. In this section we give a short review of the
results of the perturbation
series approach and we give the generalization to ordinary $D-$dimensional anti
de Sitter space.

The $2+1$ dimensional black hole anti de Sitter spacetime, recently found by
Banados et. al. \cite{ban1}, is a two-parameter family (mass $M$ and angular
momentum $J$) of solutions to the Einstein equations with a cosmological
constant $\Lambda=-1/l^2$:
\begin{equation}
ds^2=(M-\frac{r^2}{l^2})dt^2+(\frac{r^2}{l^2}-M+\frac{J^2}{4r^2})^{-1}dr^2
-Jdtd\phi+r^2d\phi^2.
\end{equation}
The solution can be obtained by a discrete identification of points in anti
de Sitter space \cite{ban2}
and thus has the local geometry of anti de Sitter space.
Globally, however, it is very different. It has two horizons $r_\pm$ and
a static limit $r_{\mbox{erg}}$ defining an ergosphere:
\begin{equation}
r_\pm=\sqrt{\frac{Ml^2}{2}\pm\frac{l}{2}\sqrt{M^2l^2-J^2}},\;\;\;\;\;\;\;r_
{\mbox{erg}}=\sqrt{M}l.
\end{equation}
Ordinary $D-$dimensional anti de Sitter space can formally be obtained by
taking $M=-1$ and $J=0$ in the line element (2.1), and by adding $D-3$ extra
angular coordinates. In that case, of course, there are no horizons and no
static limit.

The string equations of motion and constraints, in the conformal gauge, take
the form:
\begin{equation}
\ddot{x}^\mu-x''^\mu+\Gamma^\mu_{\rho\sigma}(\dot{x}^\rho\dot{x}^\sigma-
x'\rho x'^\sigma)=0,
\end{equation}
\begin{equation}
g_{\mu\nu}\dot{x}^\mu x'^\nu=g_{\mu\nu}(\dot{x}^\mu\dot{x}^\nu+
x'^\mu x'^\nu)=0,
\end{equation}
where dot and prime stand for derivatives with respect to $\tau$ and $\sigma,$
respectively.
In the string perturbation series
approach \cite{san1}, solutions to this set of equations
are obtained by expanding around an exact solution, typically taken as
the string center of mass:
\begin{equation}
x^\mu(\tau,\sigma)=q^\mu(\tau)+\eta^\mu(\tau,\sigma)+\xi^\mu(\tau,\sigma)+...
\end{equation}
where:
\begin{equation}
\ddot{q}^\mu+\Gamma^\mu_{\rho\sigma}(q)\dot{q}^\rho\dot{q}^
\sigma=0,\;\;\;\;\;\;\;g_
{\mu\nu}(q)\dot{q}^\mu\dot{q}^\nu=-m^2\alpha'^2.
\end{equation}
For the $2+1$ dimensional black hole anti de Sitter spacetime these equations
can be easily separated and integrated in terms of elementary functions
\cite{all1}. Concerning the first order perturbations $\eta^\mu(\tau,\sigma)$
it
is convenient to consider from the beginning only those
perturbations transverse to
the geodesic of the center of mass \cite{all1}. We
therefore introduce $D-1$ normal vectors
$n^\mu_R$:
\begin{equation}
g_{\mu\nu}(q)n^\mu_R\dot{q}^\nu=0,\;\;\;\;\;\;\;g_{\mu\nu}
(q)n^\mu_R n^\nu_S=\delta_{RS},
\end{equation}
and consider first order perturbations in the form:
\begin{equation}
\eta^\mu=\delta x^R n^\mu_R,
\end{equation}
where $\delta x^R$ are the comoving perturbations, i.e. the perturbations as
seen by an observer travelling with the center of mass of the string. In the
case of the $2+1$ dimensional black hole anti de Sitter spacetime it has
been shown that \cite{all1}:
\begin{equation}
\delta x^R(\tau,\sigma)=\sum_n[A_n^R e^{-i(n\sigma+\omega_n\tau)}+
\tilde{A}_n^R e^{-i(n\sigma-\omega_n\tau)}],
\end{equation}
where:
\begin{equation}
\omega_n=\sqrt{n^2+\frac{m^2\alpha'^2}{l^2}},
\end{equation}
and here:
\begin{equation}
A_n^R=(\tilde{A}_{-n}^R)^{\dag}.
\end{equation}
Notice that in de Sitter spacetime \cite{san1} the perturbations satisfy
Eq.(2.9) but with frequency $\omega_n=\sqrt{n^2-m^2\alpha'^2/l^2},$ thus
unstable modes (for $\mid n\mid<m\alpha'/l$) appear and the perturbations
blow up.
On the contrary, in the present case the
first order perturbations are completely regular and finite trigonometric
functions
oscillating with real frequencies $\omega_n$ for all $n.$

The second order perturbations $\xi^\mu(\tau,\sigma)$
are somewhat more complicated since they
couple to the first order perturbations. The explicit expressions can be found
in Ref.\cite{all1} so we shall not go into any detail here. From the center of
mass solution and the first and second order perturbations we can finally
calculate the world-sheet energy-momentum tensor:
\begin{equation}
T_{\pm\pm}=g_{\mu\nu}\partial_\pm x^\mu\partial_\pm x^\nu,
\end{equation}
where $\partial_\pm=\frac{1}{2}(\partial_\tau\pm\partial_\sigma),$ as well as
the conformal generators $L_n,\;\tilde{L}_n$:
\begin{equation}
T_{--}=\frac{1}{2\pi}\sum_n \tilde{L}_n e^{-in(\sigma-\tau)},\;\;\;
\;\;\;\;T_{++}=
\frac{1}{2\pi}\sum_n L_n e^{-in(\sigma+\tau)}.
\end{equation}
For the $2+1$ dimensional black hole anti de Sitter spacetime,
$L_0,\;\tilde{L}_0$ are given by \cite{all1}):
\begin{equation}
L_0=\pi\sum_n(\omega_n+n)^2\sum_{R=1}^{2}A_n^R\tilde{A}_{-n}^R-\frac{\pi}{2}
m^2\alpha'^2,
\end{equation}
\begin{equation}
\tilde{L}_0=\pi\sum_n(\omega_n-n)^2
\sum_{R=1}^{2}A_n^R\tilde{A}_{-n}^R-\frac{\pi}{2}m^2\alpha'^2.
\end{equation}
The classical constraints $L_0=\tilde{L}_0=0$ then provide a
formula for the mass:
\begin{equation}
m^2\alpha'^2=2\sum_n(2n^2+\frac{m^2\alpha'^2}{l^2})\sum_{R=1}^{2}A_n^R
\tilde{A}_{-n}^R,
\end{equation}
where the constants $A_n^R$ and $\tilde{A}_{-n}^R$ are subject to the
restriction:
\begin{equation}
\sum_n n\omega_n\sum_{R=1}^{2}A_n^R\tilde{A}_{-n}^R=0.
\end{equation}
Notice that in our normalization, Eq.(2.9), the
constants $A_n^R$ and $\tilde{A}_{-n}^R$ have the dimension of
length
and that Eq.(2.17)
is just the usual closed string restriction implying that there
must be an equal amount of left and right movers.

The conformal generators for $n\neq 0$ involve only the free oscillators
introduced in the second order
perturbations \cite{all1} and therefore do not lead to any further restrictions
on $A_n^R$ and $\tilde{A}_{-n}^R.$ This is in agreement with the result found
in de Sitter space \cite{san1} but is very different from Minkowski space where
there is still one un-used constraint that in principle eliminates a pair
of oscillators $A_n^R$ and $\tilde{A}_{-n}^R$ (c.f. light-cone gauge
quantization)

This concludes the discussion of the
classical picture of string propagation in the
$2+1$ dimensional black hole anti de Sitter spacetime using the string
perturbation approach.
These results can be easily generalized to
$D-$dimensional anti de Sitter space. The first order perturbations
(2.9)-(2.11) and the conformal generators (2.14)-(2.15) are independent of
the black hole mass $M$ so the results in $D-$dimensional anti de Sitter
space are obtained by simply changing the range of the index $R$ from $1,2$
to $1,2,...,(D-1),$ i.e.:
\begin{equation}
m^2\alpha'^2=2\sum_n(2n^2+\frac{m^2\alpha'^2}{l^2})\sum_{R=1}^{D-1}A_n^R
\tilde{A}_{-n}^R,
\end{equation}
\section{Quantization in Anti de Sitter Space}
\setcounter{equation}{0}
In this section we perform the canonical quantization of the closed bosonic
string in $D-$dimensional anti de Sitter space using the results of Section 2.
The first order perturbations (2.9) correspond to the action:
\begin{equation}
S^{(2)}=\frac{-1}{2\pi\alpha'}\int d\tau d\sigma\sum_{R=1}^{D-1}
[\eta^{ab}(\delta x^R)_{,a}(\delta x_R)_{,b}+\frac{m^2\alpha'^2}{l^2}\delta x^R
\delta x_R].
\end{equation}
The momentum conjugate to $\delta x^R$ is:
\begin{equation}
\Pi_R\equiv\frac{\partial S^{(2)}}{\partial(\delta x^R)_{,\tau}}=\frac{1}{\pi
\alpha'}(\delta x_R)_{,\tau},
\end{equation}
and then the canonical commutation relations become:
\begin{eqnarray}
&[\delta x^R,\;\delta x^S]=[\Pi_R,\;\Pi_S]=0,&\nonumber
\end{eqnarray}
\begin{eqnarray}
&[\Pi_R,\;\delta x^S]=-i\delta^S_R\delta(\sigma-\sigma').&
\end{eqnarray}
The constants $A^R_n$ and $\tilde{A}^R_n$ introduced in Eqs.(2.9)-(2.11),
which are now considered as quantum operators,
do not represent independent left
and right movers since the frequency $\omega_n$ is always positive. It
is therefore convenient to make the redefinitions:
\begin{eqnarray}
&A^R_n\equiv\left\{ \begin{array}{cl} \tilde{a}^R_n & n>0 \\ a^R_{-n}
& n<0 \end{array}\right.&\nonumber
\end{eqnarray}
\begin{eqnarray}
&\tilde{A}^R_n\equiv\left\{ \begin{array}{cl} a^R_{-n} & n>0 \\
\tilde{a}^R_n & n<0 \end{array}\right.&
\end{eqnarray}
so that the first order perturbations take the form:
\begin{equation}
\delta x^R=\sum_{n\neq 0}[a^R_n e^{-in(\Omega_n\tau-\sigma)}+
\tilde{a}^R_n e^{-in(\Omega_n\tau+
\sigma)}]+A^R_0 e^{-i\frac{m\alpha'}{l}\tau}+\tilde{A}^R_0
e^{i\frac{m\alpha'}{l}\tau},
\end{equation}
where:
\begin{equation}
\Omega_n=\sqrt{1+\frac{m^2\alpha'^2}{n^2 l^2}},\;\;\;\;a^R_n=
(a^R_{-n})^{\dag},\;\;\;\;\tilde{a}^R_n=(\tilde{a}^R_{-n})^{\dag},\;
\;\;\tilde{A}^R_0
=(A^R_0)^{\dag}.
\end{equation}
The non-vanishing commutators now take the form:
\begin{equation}
[a^R_n,\;(a^R_n)^{\dag}]=[\tilde{a}^R_n,\;(\tilde{a}^R_n)^{\dag}]=\frac
{\alpha'}{2n\Omega_n},\;\;\;\;\;\;[A^R_0,\;(A^R_0)^{\dag}]=\frac
{l}{2m}.
\end{equation}
and the mass formula (2.18), with symmetric order of the operators, becomes:
\begin{equation}
m^2\alpha'^2=\sum_{n>0}(2n^2+\frac{m^2\alpha'^2}{l^2})
\sum_{R=1}^{D-1}[(a^R_n)^{\dag}a^R_n+a^R_n(a^R_n)^{\dag}+
(\tilde{a}^R_n)^{\dag}\tilde{a}^R_n+\tilde{a}^R_n(\tilde{a}^R_n)^{\dag}],
\end{equation}
where we have eliminated the zero-modes.
We finally introduce the more conventionally normalized oscillators:
\begin{equation}
a^R_n=\sqrt{\frac{\alpha'}{2n\Omega_n}}\alpha^R_n,\;\;\;\;\tilde{a}^R_n=
\sqrt{\frac{\alpha'}{2n\Omega_n}}\tilde{\alpha}^R_n,
\;\;\;\;\;\;\mbox{for all}\;\;n>0
\end{equation}
The non-vanishing commutators are then represented by:
\begin{equation}
[\alpha^R_n,\;(\alpha^R_n)^{\dag}]=[\tilde{\alpha}^R_n,\;(\tilde{\alpha}^R_n)^
{\dag}]=1,\;\;\;\;\;\;\mbox{for all}\;\;n>0
\end{equation}
The classical constraints $L_0=\tilde{L}_0=0$ in the quantum theory take
the form:
\begin{equation}
(L_0-2\pi\alpha' a)\mid\psi>=(\tilde{L}_0-2\pi\alpha' a)\mid\psi>=0,
\end{equation}
where $a$ is the normal-ordering constant and the factor $2\pi\alpha'$ is
introduced for later convenience. The normal-ordering constant is most
easily obtained by symmetrization of the oscillator products as in Eq.(3.8).
Using Eqs.(2.14)-(2.15),
the physical state conditions (3.11), in terms of the conventionally
normalized oscillators, become:
\begin{equation}
m^2\alpha'=(D-1)\sum_{n>0}\frac{2n^2+m^2\alpha'\lambda}
{\sqrt{n^2+m^2\alpha'\lambda}}+\sum_{n>0}\frac{2n^2+m^2\alpha'\lambda}
{\sqrt{n^2+m^2\alpha'\lambda}}\sum_{R=1}^{D-1}[(\alpha^R_n)^{\dag}
\alpha^R_n+(\tilde{\alpha}^R_n)^{\dag}\tilde{\alpha}^R_n],
\end{equation}
and:
\begin{equation}
\sum_{n>0} n\sum_{R=1}^{D-1}[(\alpha^R_n)^{\dag}
\alpha^R_n-(\tilde{\alpha}^R_n)^{\dag}\tilde{\alpha}^R_n]=0,
\end{equation}
where we introduced the dimensionless positive
parameter $\lambda=\alpha'/l^2.$
The first term in the mass formula (3.12)
obviously needs zeta-function regularization. Assuming $\lambda<<1,$ which
is clearly fulfilled in most interesting cases, we find for the lower mass
states $m^2\alpha'\lambda<<1$:
\begin{equation}
\sum_{n>0}\frac{2n^2+m^2\alpha'\lambda}
{\sqrt{n^2+m^2\alpha'\lambda}}=-\frac{1}{6}+\frac{(m^2\alpha')^2}
{4}\zeta (3)\lambda^2-\frac{(m^2\alpha')^3}{4}\zeta (5)\lambda^3+
{\cal O}((m^2\alpha'\lambda)^4).
\end{equation}
For the very high mass states $m^2\alpha'\lambda>>1$ (but still assuming
$\lambda<<1$) we find instead:
\begin{equation}
\sum_{n>0}\frac{2n^2+m^2\alpha'\lambda}
{\sqrt{n^2+m^2\alpha'\lambda}}=\frac{m^2\alpha'\lambda}{2}-\frac{1}{2}
\sqrt{m^2\alpha'\lambda}+{\cal O}(1).
\end{equation}
In de Sitter space the mass formula takes the form (3.12) but with $\lambda$
being negative \cite{san1}. This
means that there is only a {\it finite} number of
states in de Sitter space. Beyond some
maximal mass $m^2\alpha'\sim 1/\lambda,$
the strings become unstable and no real solutions
to the analogue of Eq.(3.12) can be found. Here in anti de Sitter
space $\lambda$ is positive so that arbitrarily high mass solutions of
Eq.(3.12) can be found. We therefore find {\it infinitely} many states in
anti de Sitter space.

Let us first consider the lower mass spectrum in a little more detail.
As in Minkowski space, is convenient to characterize the physical
states by the
eigenvalue of the number operator:
\begin{equation}
N=\frac{1}{2}\sum_{n>0} n\sum_{R=1}^{D-1}[(\alpha^R_n)^{\dag}
\alpha^R_n+(\tilde{\alpha}^R_n)^{\dag}\tilde{\alpha}^R_n].
\end{equation}
\vskip 6pt
\hspace*{-6mm}For $N=0$ we have the vacuum state $\mid 0>$ defined by:
\begin{equation}
\alpha^R_n\mid 0>=\tilde{\alpha}^R_n\mid 0>=0,
\;\;\;\;\;\;\mbox{for all}\;\;n>0.
\end{equation}
Using Eqs.(3.12) and (3.14) we find that there is a tachyon with mass:
\begin{equation}
<0\mid m^2\alpha'\mid 0>=(D-1)
[-\frac{1}{6}+\frac{(D-1)^2}{144}\zeta(3)\lambda^2+
\frac{(D-1)^3}{864}\zeta(5)\lambda^3+{\cal O}(\lambda^4)].
\end{equation}
\vskip 6pt
\hspace*{-6mm}At the first excited level ($N=1$) we have states of the form:
\begin{equation}
(\tilde{\alpha}^R_1)^{\dag}(\alpha^S_1)^{\dag}\mid 0>\equiv\mid
\Omega_{1\hspace*{1mm}1}^{\tilde{R}S}>.
\end{equation}
For $D=25$ it yields the graviton:
\begin{equation}
<\Omega_{1\hspace*{1mm}1}^{\tilde{R}S}\mid m^2\alpha'\mid
\Omega_{1\hspace*{1mm}1}^{\tilde{R}S}>=0.
\end{equation}
Notice that $m^2\alpha'=0$ is an {\it exact} solution of the mass formula in
$D=25$ dimensions. As in de Sitter space \cite{san1}, this indicates that the
critical dimension in
anti de Sitter space is 25. It should be stressed, however, that it is not
known how to obtain de Sitter space from the
$\beta$-function equations. In the further analysis we take $D=25.$
\vskip 6pt
\hspace*{-6mm}At the next excited level ($N=2$) we have states of the form:
\begin{eqnarray}
&(\tilde{\alpha}^R_1)^{\dag}({\alpha}^S_1)^{\dag}(\tilde{\alpha}^T_1)^{\dag}
({\alpha}^U_1)^{\dag}\mid 0>\equiv\mid
\Omega_{1\hspace*{1mm}1\hspace*{1mm}1\hspace*{1mm}1}^{\tilde{R}
S\tilde{T}U}>&\nonumber
\end{eqnarray}
\begin{eqnarray}
&(\tilde{\alpha}^R_1)^{\dag}(\tilde{\alpha}^S_1)^{\dag}({\alpha}^T_2)^{\dag}
\mid 0>\equiv\mid\Omega_{1\hspace*{1mm}1\hspace*{1mm}2}^
{\tilde{R}\tilde{S}T}>,\;\;\;\;
({\alpha}^R_1)^{\dag}({\alpha}^S_1)^{\dag}(\tilde{\alpha}^T_2)^{\dag}
\mid 0>\equiv\mid\Omega_{1\hspace*{1mm}1\hspace*{1mm}2}^
{{R}{S}\tilde{T}}>&\nonumber
\end{eqnarray}
\begin{eqnarray}
&(\tilde{\alpha}^R_2)^{\dag}(\alpha^S_2)^{\dag}\mid 0>\equiv\mid
\Omega_{2\hspace*{1mm}2}^{\tilde{R}S}>.&
\end{eqnarray}
In Minkowski spacetime the
corresponding states all have $m^2\alpha'=4.$ In fact,
the mass operator and the number operator are related by $m^2\alpha'=-4+4N$ in
Minkowski space. Here in anti de Sitter space there is no such simple relation.
Using Eqs.(3.12) and (3.14) we find the following masses of the states (3.21)
when $D=25$:
\begin{eqnarray}
\hspace*{-5mm}&<&\hspace*{-2mm}\Omega_{1\hspace*{1mm}1
\hspace*{1mm}1\hspace*{1mm}1}^
{\tilde{R}S\tilde{T}U}\mid m^2\alpha'\mid
\Omega_{1\hspace*{1mm}1\hspace*{1mm}1\hspace*{1mm}1}^
{\tilde{R}S\tilde{T}U}>=4+(16+96\zeta(3))\lambda^2-
(64+384\zeta(5))\lambda^3+{\cal O}(\lambda^4)\nonumber
\end{eqnarray}
\begin{eqnarray}
\hspace*{-5mm}&<&\hspace*{-2mm}\Omega_{1\hspace*{1mm}1\hspace*{1mm}2}^
{\tilde{R}\tilde{S}T}\mid m^2\alpha'\mid
\Omega_{1\hspace*{1mm}1\hspace*{1mm}2}^
{\tilde{R}\tilde{S}T}>=<\Omega_{1\hspace*{1mm}1\hspace*{1mm}2}^
{{R}{S}\tilde{T}}\mid
m^2\alpha'\mid \Omega_{1\hspace*{1mm}1\hspace*{1mm}2}^
{{R}{S}\tilde{T}}>=\nonumber\\
\hspace*{-3mm}& &\hspace*{-1mm}\hspace*{1cm}=4+(17/2+96\zeta(3))\lambda^2-
(65/2+384\zeta(5))\lambda^3+{\cal O}(\lambda^4)
\end{eqnarray}
\begin{eqnarray}
\hspace*{-5mm}&<&\hspace*{-2mm}\Omega_{2\hspace*{1mm}2}^
{\tilde{R}S} \mid m^2\alpha'\mid
\Omega_{2\hspace*{1mm}2}^{\tilde{R}S}>=
4+(1+96\zeta(3))\lambda^2-(1+384\zeta(5))\lambda^3+{\cal O}
(\lambda^4)\nonumber
\end{eqnarray}
We therefore reach the interesting conclusion that the coupling to the
gravitational background (here anti de Sitter spacetime) gives rise to a
{\it fine structure} in the string mass spectrum. This turns out to be a
general feature at all excited levels beyond the graviton, i.e. for $N>1.$
To zeroth order in the expansion parameter $\lambda$ we recover, of course,
the flat Minkowski space spectrum. The first corrections appear to order
$\lambda^2.$ The leading Regge trajectory for the lower mass states
therefore takes the form:
\begin{equation}
J=2+\frac{1}{2}m^2\alpha'+{\cal O}(\lambda^2).
\end{equation}
To second order in $\lambda$ the masses are the same as in de Sitter space
\cite{san1}. The difference in the lower mass spectrum between de Sitter space
and anti de Sitter space is of order $\lambda^3.$

For the very high mass
spectrum the situation changes drastically. Consider first excited states
of the form:
\begin{equation}
(\tilde{\alpha}^{R_1}_1)^{\dag}(\alpha^{S_1}_1)^{\dag},.\;.\;.\;.\;.\;.,
(\tilde{\alpha}^{R_N}_1)^{\dag}(\alpha^{S_N}_1)^{\dag}\mid 0>
\equiv\mid\Omega_{1\hspace*{3mm}1\;....\;1\hspace*{3mm}1}^
{\tilde{R}_1 S_1....\tilde{R}_N S_N}>.
\end{equation}
This is a state with eigenvalue $N$ of the number operator, and we will
consider very large $N,$ say $N>>\lambda^{-1}.$ Using Eqs.(3.12) and
(3.15) we find the approximate value of the mass:
\begin{equation}
<\Omega_{1\hspace*{3mm}1\;....\;1\hspace*{3mm}1}^
{\tilde{R}_1 S_1....\tilde{R}_N S_N}\mid
m^2\alpha'\mid \Omega_{1\hspace*{3mm}1\;....\;1\hspace*{3mm}1}^{\tilde{R}_1 S_1
....\tilde{R}_N S_N}>\approx 4\lambda N^2
\end{equation}
The Regge trajectory now takes the form:
\begin{equation}
J^2\approx
\frac{1}{\lambda}m^2\alpha',\;\;\;\;\;\;\mbox{for}\;\;N>>\lambda^{-1}
\end{equation}
This is significantly different from the lower mass relation, as compared with
Eq.(3.23). Considering instead the state (for even $N$):
\begin{equation}
(\tilde{\alpha}^{R_1}_2)^{\dag}(\alpha^{S_1}_2)^{\dag},.\;.\;.\;.\;.\;.,
(\tilde{\alpha}^{R_{N/2}}_2)^{\dag}(\alpha^{S_{N/2}}_2)^{\dag}\mid 0>
\equiv\mid\Omega_{2\hspace*{2mm}2\;\;....\;\;2\hspace*{6mm}2}^{\tilde{R}_1 S_1
....\tilde{R}_{N/2} S_{N/2}}>,
\end{equation}
and taking again $N>>\lambda^{-1},$ we find:
\begin{equation}
<\Omega_{2\hspace*{2mm}2\;\;....\;\;2\hspace*{6mm}2}^{\tilde{R}_1 S_1
....\tilde{R}_{N/2} S_{N/2}}\mid m^2\alpha'\mid
\Omega_{2\hspace*{2mm}2\;\;....\;\;2\hspace*{6mm}2}^{\tilde{R}_1 S_1
....\tilde{R}_{N/2} S_{N/2}}>\approx \lambda N^2
\end{equation}
In flat Minkowski
space the states Eq.(3.24) and Eq.(3.27) have the same mass ($m^2\alpha'=
-4+4N\approx 4N$) but here in anti de Sitter space the masses are
different by a factor of $4.$ The fine structure we found in the lower
mass spectrum completely changes the very high mass spectrum. States with the
same eigenvalue of the number operator can have considerably different
masses. Furthermore, states with different eigenvalues
of the number operator can get mixed up in the mass spectrum.
For very high mass states of the form Eq.(3.24) or Eq.(3.27), the level spacing
($\Delta(m^2\alpha')$ as a function of $N$) grows
proportionally to $N.$ This should be contrasted with the situation in
Minkowski space where the level spacing is constant. This suggests that in
anti de Sitter space the partition function can be defined for any temperature,
as opposed to Minkowski space where there is a critical temperature (the
Hagedorn temperature) of the
order $(\alpha')^{-1/2}.$ In Minkowski space, for very large $N,$
the number of states $d_N$ with eigenvalue $N$ of
the number operator is roughly growing like \cite{green}:
\begin{equation}
d_N\sim \frac{e^{\sqrt{N}}}{N^p},
\end{equation}
where $p$ is some number larger than $1.$ This
holds for anti de Sitter space as well. In anti de Sitter space, for
the states
in the form (3.24) or (3.27), this leads to the following
density of levels as a function of mass:
\begin{equation}
\rho(m)\sim \frac{e^{\beta_l\sqrt{m}}}{m^{p-1}},
\end{equation}
where $\beta_l\sim\sqrt{l}\;$ is {\it independent} of the string tension.
Therefore, for a gas of strings in anti de Sitter space at temperature
$\beta^{-1},$ the partition function behaves like:
\begin{eqnarray}
Z(\beta)&=&\int^\infty dm\;\rho(m)e^{-\beta m}\nonumber\\
&\sim &\int^\infty dm\;\frac{e^{\beta_l\sqrt{m}(1-\frac{\beta}{\beta_l}
\sqrt{m})}}{m^{p-1}}.
\end{eqnarray}
This integral is finite for any value of $\beta,$ thus we find no Hagedorn
temperature in anti de Sitter space.

We close this section with some interesting remarks on the string masses in the
two regimes considered. For the low mass states ($m^2\alpha'\lambda<<1$) our
results can be written as:
\begin{equation}
<j\mid m^2(\alpha',l)\mid j>=\frac{4(N-1)}{\alpha'}+\frac{1}{l^2}
\sum_{n=0}^{\infty}a_{jn}(\alpha'/l^2)^n,
\end{equation}
where "$j$" is a collective index labelling the state $\mid j>.$ It is now
important to notice that $a_{j0}=0$ for {\it all} the low mass states, i.e.
there is no "constant" term on the right hand side of Eq.(3.32). A non-zero
$a_{j0}$-term would give rise to a $\alpha'$-independent contribution to
the string mass. Its absense, on the other hand, means that the first term
on the right hand side of Eq.(3.32) is super-dominant (since, in all cases,
$\alpha'/l^2=\lambda<<1$) and that the string scale is therefore set by
$1/\alpha'.$ For the high mass states ($m^2\alpha'\lambda>>1$) we found
instead:
\begin{equation}
<j\mid m^2(\alpha',l)\mid j>\approx\frac{d_j}{l^2}N^2,\;\;\;\;\;\;\mbox{for}\;
N>>l^2/\alpha'
\end{equation}
where the number $d_j$ depends on the state. The masses of the high mass
states are therefore {\it independent} of $\alpha'.$ Moreover,
the right hand side of Eq.(3.33) is exactly like a non-zero dominant
$a_{j0}$-term in Eq.(3.32). For the high mass states the scale is therefore
set by $1/l^2$ which is equal to the absolute value of the cosmological
constant $\Lambda$ (up to a geometrical factor). This suggests that for
$\lambda<<1,$ the masses of {\it all} string states can be represented by a
formula of the form (3.32). For the low mass states $a_{j0}=0,$ while for
the high mass states $a_{j0}$ becomes a large positive number.
\section{The Black String Background}
\setcounter{equation}{0}
By a duality transformation \cite{wel} the $2+1$ dimensional black hole anti
de Sitter spacetime becomes the black string of Horne and Horowitz
\cite{hor}. It
is therefore interesting to compare the string propagation in these two
spacetimes. In Section 2 we
presented the main results of the string perturbation
approach in the case of the $2+1$ dimensional black hole anti de Sitter
spacetime; the details can be found in Ref.\cite{all1}. In Ref.\cite{all1} we
considered also the first order string perturbations in the black string
background, with special interest in the behaviour near the physical
singularity. In this section we perform a more complete analysis of the
string propagation in the black string background. We consider all possible
string center of mass geodesics (bounded and unbounded) and we calculate
explicitly the first and second order string perturbations around these
solutions. In the asymptotic
region we then calculate the world-sheet energy-momentum tensor and we derive
the mass formula of the string. To allow
for comparison with the results of Section 1 we
consider the uncharged black string with line element \cite{hor}:
\begin{equation}
ds^2=-(1-\frac{Ml}{r})dt^2+(1-\frac{Ml}{r})^{-1}\frac{l^2dr^2}{4r^2}+dx^2.
\end{equation}
In this form the spacetime is just the product of Witten's 2 dimensional
black hole \cite{wit} and the real line space. It has a horizon at $r=Ml$
and, contrary to its dual the $2+1$ dimensional black hole anti de Sitter
spacetime, it has a strong curvature singularity at $r=0.$  For a radially
infalling string ($x=$const.) the geodesic equations (2.6), determining
the string center of mass motion, are integrated to:
\begin{equation}
\dot{r}^2=\frac{4r^2\alpha'^2}{l^2}[E^2-m^2+\frac{m^2Ml}{r}],
\end{equation}
\begin{equation}
\dot{t}=\frac{E\alpha'}{1-Ml/r},
\end{equation}
where $E$ is the integration constant. These equations can be solved in terms
of elementary functions:
\begin{equation}
r(\tau)=\frac{m^2Ml}{E^2-m^2}\sinh^2(\frac{\alpha'\sqrt{E^2-m^2}}{l}\tau),
\end{equation}
\begin{equation}
t(\tau)=E\alpha'\tau+l\;\mbox{arcth}[\frac{E}{\sqrt{E^2-m^2}}\mbox{th}
(\frac{\alpha'\sqrt{E^2-m^2}}{l}\tau)].
\end{equation}
Notice that these relations are well defined for any relation between
$E^2$ and $m^2.$ For $E^2-m^2<0\;,$ $r(\tau)$ becomes a trigonometric
function describing the bounded solutions. For $E^2=m^2$ we find that $r(\tau)$
is simply proportional to $\tau^2,\;\;r(\tau)=M(m\alpha'\tau)^2/l.$

Two covariantly constant normal vectors fulfilling Eqs.(2.7) are given by:
\begin{equation}
n^\mu_\perp=(0,\;0,\;1),\;\;\;\;\;\;n^\mu_\parallel=(\frac{l\dot{r}}
{2m\alpha'(r-Ml)},\;\frac{2Er}{ml},\;0).
\end{equation}
It can now be shown \cite{all1} that the first order perturbations (2.8)
take the form:
\begin{equation}
\delta x^\perp(\tau,\sigma)=\sum_n C_{n}^\perp(\tau)e^{-in\sigma},\;\;\;\;\;\;
\delta x^\parallel(\tau,\sigma)=\sum_n C_{n}^\parallel(\tau)e^{-in\sigma},
\end{equation}
where $C_{n}^\perp$ and $C_{n}^\parallel$ are solutions of the
"Schr\"{o}dinger equations" in $\tau$:
\begin{equation}
\ddot{C}_{n}^\perp+n^2 C_{n}^\perp=0,
\end{equation}
\begin{equation}
\ddot{C}_{n}^\parallel+(n^2-\frac{2Mm^2\alpha'^2}{lr})C_{n}^\parallel=0.
\end{equation}
Not surprisingly the perturbations in the "transverse" direction are completely
finite and regular trigonometric functions:
\begin{equation}
C_n^\perp(\tau)=A_n^\perp e^{-in\tau}+B_n^\perp e^{in\tau},
\end{equation}
where $A_n^\perp=(A_{-n}^\perp)^{\dag},\;\;B_n^\perp=(B_{-n}^\perp)^{\dag}.$ In
Ref.\cite{all1} we considered the solution of Eq.(4.9) for $r\rightarrow 0$ and
we found that the perturbations blow up. We now
give the complete solution in explicit form for all $r.$
\vskip 6pt
\hspace*{-6mm}For $E^2=m^2,$ using Eq.(4.4) we find:
\begin{equation}
\ddot{C}_{n}^\parallel+(n^2-\frac{2}{\tau^2})C_{n}^\parallel=0.
\end{equation}
This equation is solved by:
\begin{equation}
C_n^\parallel(\tau)=A_n^\parallel(1-\frac{i}{n\tau})e^{-in\tau}+
B_n^\parallel(1+\frac{i}{n\tau})e^{in\tau},
\end{equation}
where $A_n^\parallel=(A_{-n}^\parallel)^{\dag},\;\;
B_n^\parallel=(B_{-n}^\parallel)^{\dag}.$ Notice in particular the
following behaviour:
\begin{equation}
C_n^\parallel(\tau)\rightarrow\left\{ \begin{array}{ll} A_n^\parallel e^
{-in\tau}+B_n^\parallel e^{in\tau}, & \mbox{for}\;\;\tau\rightarrow-\infty\;\;
(r\rightarrow\infty) \\
\frac{i}{n\tau}(B_n^\parallel-A_n^\parallel), & \mbox{for}\;\;
\tau\rightarrow 0_-\;\;
(r\rightarrow 0) \end{array}\right.
\end{equation}
Asymptotically this is a plane wave while near the singularity the
perturbations
blow up. This in agreement with the results obtained in Ref.\cite{all1}.
\vskip 6pt
\hspace*{-6mm}For $E^2\neq m^2$ we introduce a real parameter $z$ and a
function $g(z)$:
\begin{equation}
z\equiv-\sinh^2(\frac{\alpha'\sqrt{E^2-m^2}}{l}\tau),\;\;\;\;\;\;g(z)\equiv
\frac{C_n^\parallel(z)}{z}.
\end{equation}
Eq.(4.9) then reduces to the Hypergeometric equation \cite{abr}:
\begin{equation}
z(z-1)\frac{d^2g(z)}{dz^2}+[c-(a+b+1)z]\frac{dg(z)}{dz}-ab\;g(z)=0,
\end{equation}
with parameters:
\begin{eqnarray}
a\hspace*{-2mm}&=&\hspace*{-2mm}1+\frac{inl}{2\alpha'\sqrt{E^2-m^2}}\nonumber\\
b\hspace*{-2mm}&=&\hspace*{-2mm}1-\frac{inl}{2\alpha'\sqrt{E^2-m^2}}\\
c\hspace*{-2mm}&=&\hspace*{-2mm}\frac{5}{2}\nonumber
\end{eqnarray}
For $\mid z\mid\leq 1$ the solution $C_n^\parallel(z)$
is a linear combination of the functions:
\begin{equation}
zF(a,b,c;z)\;\;\;\;\mbox{and}\;\;\;\;z^{2-c}F(a-c+1,b-c+1,2-c;z)
\end{equation}
In the case where $E^2<m^2$ (where $r(\tau)$ reduces to a trigonometric
function) we find that $z\in[0,\;1]$ and therefore Eq.(4.17)
gives the full solution.
For $E^2>m^2,$ on the other hand, we have $z\in\;]-\infty,\;0]$ and the
solutions (4.17) have to be matched with solutions for $\mid z\mid>1$ using
analytical continuation \cite{abr}. This leads to the following
expression for $C_n^\parallel(z)$ when $\mid z\mid>1:$
\begin{eqnarray}
C_n^\parallel(z)\hspace*{-2mm}&=&\hspace*{-2mm}2^{\frac{inl}{\alpha'
\sqrt{E^2-m^2}}}A_n^\parallel\;(-z)^{1-b}F(b,1-c+b,1-a+b;1/z)\nonumber\\
\hspace*{-2mm}&+&\hspace*{-2mm}2^{\frac{-inl}{\alpha'
\sqrt{E^2-m^2}}}B_n^\parallel\;(-z)^{1-a}F(a,1-c+a,1-b+a;1/z)
\end{eqnarray}
where again $A_n^\parallel=(A_{-n}^\parallel)^{\dag},\;\;
B_n^\parallel=(B_{-n}^\parallel)^{\dag}.$
The constant factors
in front of $A_n^\parallel$ and $B_n^\parallel$ were included
to ensure the asymptotic behaviour:
\begin{equation}
C_n^\parallel(\tau)\rightarrow A_n^\parallel e^
{-in\tau}+B_n^\parallel e^{in\tau},\;\;\mbox{for}\;\;\tau\rightarrow
-\infty\;\;
(r\rightarrow\infty).
\end{equation}
In terms of the constants $A_n^\parallel$ and $B_n^\parallel$ the solution
for $\mid z\mid\leq 1$ reads:
\begin{eqnarray}
C_n^\parallel(z)=\hspace*{-2mm}&-&\hspace*{-2mm}
\frac{2nl\;2^{\frac{inl}{\alpha'\sqrt{E^2-m^2}}}}
{3\alpha'\sqrt{E^2-m^2}}A_n^\parallel[i
\frac{\Gamma(2-c)\Gamma(b-a)}{\Gamma(b-c+1)\Gamma(1-a)}zF(a,b,c;z)\nonumber\\
\hspace*{-2mm}&+&\hspace*{-2mm}\frac{\Gamma(c)\Gamma(b-a)}{\Gamma(b)\Gamma(c-a)}
z^{2-c}F(a-c+1,b-c+1,2-c;z)]\nonumber\\
\hspace*{-2mm}&+&\hspace*{-2mm}\frac
{2nl\;2^{\frac{-inl}{\alpha'\sqrt{E^2-m^2}}}}
{3\alpha'\sqrt{E^2-m^2}}B_n^\parallel[i
\frac{\Gamma(2-c)\Gamma(a-b)}{\Gamma(a-c+1)\Gamma(1-b)}zF(a,b,c;z)\nonumber\\
\hspace*{-2mm}&+&\hspace*{-2mm}\frac{\Gamma(c)\Gamma(a-b)}{\Gamma(a)
\Gamma(c-b)}z^{2-c}F(a-c+1,b-c+1,2-c;z)]
\end{eqnarray}
For $r\rightarrow 0\;\;(\tau\rightarrow 0_-)$ we find:
\begin{equation}
C_n^\parallel(\tau\rightarrow 0_-)\rightarrow
-\frac{(\alpha'\sqrt{E^2-m^2}-inl)B_n^\parallel+
(\alpha'\sqrt{E^2-m^2}+inl)A_n^\parallel}{\alpha'^2(E^2-m^2)+n^2l^2}\;\frac{l}
{\tau},
\end{equation}
i.e. the perturbations blow up, in agreement with the result found in
Ref.\cite{all1}.
\vskip 6pt
\hspace*{-6mm}The second order perturbations $\xi^\mu(\tau,\sigma)$ are
determined by \cite{san1,men}:
\begin{equation}
\dot{q}^\lambda\nabla_\lambda(\dot{q}^\delta\nabla_\delta\xi^\mu)-
R^\mu_{\epsilon\delta\lambda}\dot{q}^\epsilon\dot{q}^\delta\xi^\lambda-
\xi''^\mu=U^\mu,
\end{equation}
where the source $U^\mu$ is bilinear in the first order perturbations, and
explicitly given by:
\begin{equation}
U^\mu=-\Gamma^\mu_{\rho\sigma}(\dot{\eta}^\rho\dot{\eta}^\sigma-
\eta'^\rho\eta'^\sigma)-2\Gamma^\mu_{\rho\sigma,\lambda}\dot{q}^\rho
\eta^\lambda\dot{\eta}^\sigma-\frac{1}{2}\Gamma^\mu_{\rho\sigma,
\lambda\delta}\dot{q}^\rho\dot{q}^\sigma\eta^\lambda\eta^\delta.
\end{equation}
The $\xi^x$ perturbations decouple and Eq.(4.22) reduces to the free
wave equation. The $\xi^x$ perturbations are then given explicitly by:
\begin{equation}
\xi^x(\tau,\sigma)=\sum_n[\tilde{A}_n^x e^{-in(\tau+\sigma)}+
\tilde{B}_n^x e^{-in(\sigma-\tau)}].
\end{equation}
where $\tilde{A}_n^x=(\tilde{A}_{-n}^x)^{\dag},\;\;
\tilde{B}_n^x=(\tilde{B}_{-n}^x)^{\dag}.$
The perturbations $\xi^t$ and $\xi^r$ are somewhat more complicated to
derive. By redefining $\xi^r$ and $U^r$:
\begin{equation}
\xi^r\equiv\frac{2r}{l}(1-\frac{Ml}{r})\xi^*,
\end{equation}
\begin{equation}
U^r\equiv\frac{2r}{l}(1-\frac{Ml}{r})U^*,
\end{equation}
we find from eq. (4.22):
\begin{equation}
\left( \begin{array}{c} \ddot{\xi}^t \\  \ddot{\xi}^* \end{array}\right)-
\left( \begin{array}{c} \xi''^t \\ \xi''^* \end{array}\right)+
2{\cal A}\left( \begin{array}{c} \dot{\xi}^t \\ \dot{\xi}^* \end{array}\right)+
{\cal B}\left( \begin{array}{c} \xi^t \\ \xi^* \end{array}\right)=
\left( \begin{array}{c} U^t \\ U^* \end{array}\right),
\end{equation}
where the matrices ${\cal A}$ and ${\cal B}$ are given by:
\begin{eqnarray}
{\cal A}=\left( \begin{array}{cc} \alpha & \beta \\ \beta &
\alpha \end{array}\right), \;\;\;\;
{\cal B}=\left( \begin{array}{cc} 0 & \gamma \\ 0 &
\delta \end{array}\right);\nonumber
\end{eqnarray}
\begin{eqnarray}
\alpha=\frac{Ml\dot{r}}{2r^2}(1-\frac{Ml}{r})^{-1},\;\;\;\;\beta=
\frac{ME\alpha'}{r}(1-\frac{Ml}{r})^{-1},
\end{eqnarray}
\begin{eqnarray}
\gamma=\frac{-2EM\alpha'\dot{r}}{r^2(1-Ml/r)},\;\;\;\;
\delta=\frac{-4ME^2\alpha'^2}{lr(1-Ml/r)}+\frac{2Mm^2\alpha'^2}{lr}.\nonumber
\end{eqnarray}
The first order $\tau$-derivatives in eq. (4.27) are eliminated by the
transformation:
\begin{equation}
\left( \begin{array}{c} \xi^t \\ \xi^* \end{array}\right)\equiv
{\cal G}\left( \begin{array}{c} \Sigma^t \\ \Sigma^*
\end{array}\right);\;\;\;\;
{\cal G}=e^{-\int^\tau{\cal A}(\tau')d\tau'},
\end{equation}
i.e.:
\begin{equation}
{\cal G}=(1-Ml/r)^{-1}\left( \begin{array}{cc}
E/m & l\dot{r}/2m\alpha' r \\ l\dot{r}/2m\alpha' r & E/m \end{array}\right).
\end{equation}
We now Fourier expand the second order perturbations and the sources:
\begin{equation}
\Sigma^t(\tau,\sigma)=\sum_n\Sigma^t_n(\tau)e^{-in\sigma},\;\;\;\;\Sigma^*
(\tau,\sigma)=
\sum_n\Sigma^*_n(\tau)e^{-in\sigma},
\end{equation}
\begin{equation}
U^t(\tau,\sigma)=\sum_n U^t_n(\tau)e^{-in\sigma},\;\;\;\; U^*(\tau,\sigma)=
\sum_n U^*_n(\tau)e^{-in\sigma},
\end{equation}
and the matrix equation (4.27) reduces to:
\begin{equation}
\left( \begin{array}{c} \ddot{\Sigma}^t_n \\ \ddot{\Sigma}^*_n
\end{array}\right)+{\cal V}
\left( \begin{array}{c} \Sigma^t_n \\ \Sigma^*_n
\end{array}\right)={\cal G}^{-1}
\left( \begin{array}{c} U^t_n \\ U^*_n \end{array}\right)\equiv
\left( \begin{array}{c} \tilde{U}^t_n \\ \tilde{U}^*_n
\end{array}\right),
\end{equation}
where:
\begin{equation}
{\cal V}={\cal G}^{-1}(n^2 I+{\cal B}-{\cal A}^2-\dot{{\cal A}}){\cal G}=
\left( \begin{array}{cc} n^2 & 0 \\ 0 & n^2-\frac{2Mm^2\alpha'^2}{lr}
\end{array} \right),
\end{equation}
i.e. two decoupled inhomogeneous second order linear differential equations.
These equations can easily be integrated by noticing that the corresponding
homogeneous equations take exactly the same form as Eqs.(4.8)-(4.9), which
we have already solved explicitly. ${\Sigma}^t_n$ and $\Sigma^*_n$ will
then take the same form, plus extra terms involving suitable integrals of the
sources. By transforming backwards, using Eqs.(4.25)-(4.32), we finally get the
explicit expressions for $\xi^r$ and $\xi^t$ which we shall however
not write down here.
It is instructive to consider the results in the asymptotic region
$r\rightarrow\infty.$ This region can only be reached if $E^2\geq m^2$ and for
the following computations we therefore assume $E^2>m^2,$ although the results
will hold for $E^2=m^2,$ too. Using the results of the analysis of the
first order perturbations we find for $r\rightarrow\infty\;\;(\tau\rightarrow
-\infty):$
\begin{equation}
\eta^t(\tau,\sigma)=-\frac{\sqrt{E^2-m^2}}{m}\sum_n[{A}_n^\parallel
e^{-in(\tau+\sigma)}+
{B}_n^\parallel e^{-in(\sigma-\tau)}]+{\cal O}(1/r),
\end{equation}
\begin{equation}
\eta^r(\tau,\sigma)=\frac{2Er}{ml}\sum_n[{A}_n^\parallel
e^{-in(\tau+\sigma)}+
{B}_n^\parallel e^{-in(\sigma-\tau)}]+{\cal O}(1).
\end{equation}
The sources then take the asymptotic form:
\begin{equation}
\left( \begin{array}{c} \tilde{U}^t \\
\tilde{U}^* \end{array}\right)=
\frac{2E^2}{lm^2}[(\frac{\partial\delta x^\parallel}{\partial\tau})^2
-(\frac{\partial\delta x^\parallel}{\partial\sigma})^2]
\left( \begin{array}{c} \sqrt{E^2-m^2}/m \\
E/m \end{array}\right)+
{\cal O}(1/r),
\end{equation}
and Eqs.(4.33) are solved by:
\begin{eqnarray}
{\Sigma}^t_n(\tau)\hspace*{-2mm}&=&\hspace*{-2mm}\tilde{A}_n^\parallel
e^{-in\tau}+\tilde{B}_n^\parallel e^{in\tau}+\frac{2E^2\sqrt{E^2-m^2}}
{lm^3}\;\frac{e^{in\tau}}{n}\sum_p(n-p)A_p^\parallel B_{n-p}^\parallel
e^{-2ip\tau}\nonumber\\
\hspace*{-2mm}&+&\hspace*{-2mm}\frac{2E^2\sqrt{E^2-m^2}}
{lm^3}\;\frac{e^{-in\tau}}{n}\sum_p(n-p)A_{n-p}^\parallel B_{p}^\parallel
e^{2ip\tau}+{\cal O}(1/r)
\end{eqnarray}
\begin{eqnarray}
{\Sigma}^*_n(\tau)\hspace*{-2mm}&=&\hspace*{-2mm}\tilde{C}_n^\parallel
e^{-in\tau}+\tilde{D}_n^\parallel e^{in\tau}+\frac{2E^3}
{lm^3}\;\frac{e^{in\tau}}{n}\sum_p(n-p)A_p^\parallel B_{n-p}^\parallel
e^{-2ip\tau}\nonumber\\
\hspace*{-2mm}&+&\hspace*{-2mm}\frac{2E^3}
{lm^3}\;\frac{e^{-in\tau}}{n}\sum_p(n-p)A_{n-p}^\parallel B_{p}^\parallel
e^{2ip\tau}+{\cal O}(1/r)
\end{eqnarray}
The second order perturbations become:
\begin{equation}
\xi^t(\tau,\sigma)=\sum_n[(\frac{E}{m}\tilde{A}_n^\parallel-\frac{\sqrt
{E^2-m^2}}{m}\tilde{C}_n^\parallel)e^{-in(\tau+\sigma)}+
(\frac{E}{m}\tilde{B}_n^\parallel-\frac{\sqrt
{E^2-m^2}}{m}\tilde{D}_n^\parallel)e^{-in(\sigma-\tau)}]+{\cal O}(1/r),
\end{equation}
\begin{eqnarray}
\xi^r(\tau,\sigma)=\frac{2r}{l}\sum_n[(\frac{E}{m}
\tilde{C}_n^\parallel\hspace*{-2mm}&-&\hspace*{-2mm}\frac{\sqrt
{E^2-m^2}}{m}\tilde{A}_n^\parallel)e^{-in(\tau+\sigma)}+
(\frac{E}{m}\tilde{D}_n^\parallel-\frac{\sqrt
{E^2-m^2}}{m}\tilde{B}_n^\parallel)e^{-in(\sigma-\tau)}\nonumber\\
\hspace*{-2mm}&+&\hspace*{-2mm}\frac{2E^2}{lm^2}e^{-in(\sigma-\tau)}
\sum_p A_p^\parallel B_{n-p}^\parallel e^{-2ip\tau}]+{\cal O}(1).
\end{eqnarray}
$\eta^r$ and $\xi^r$ are ordinary plane waves in the coordinate $x^R$
defined by:
\begin{equation}
x^R=R+\eta^R+\xi^R+...\equiv \frac{l}{2}\log\frac{x^r}{l}=\frac{l}{2}\log
\frac{r+\eta^r+\xi^r+...}{l}.
\end{equation}
We find:
\begin{equation}
\eta^R(\tau,\sigma)=\frac{E}{m}\sum_n[{A}_n^\parallel
e^{-in(\tau+\sigma)}+
{B}_n^\parallel e^{-in(\sigma-\tau)}]+{\cal O}(e^{-2R/l}),
\end{equation}
\begin{eqnarray}
\xi^R(\tau,\sigma)=\sum_n[(\frac{E}{m}
\tilde{C}_n^\parallel\hspace*{-2mm}&-&\hspace*{-2mm}\frac{\sqrt
{E^2-m^2}}{m}\tilde{A}_n^\parallel-\frac{E^2}{lm^2}\sum_p
A_{n-p}^\parallel A_p^\parallel)e^{-in(\tau+\sigma)}\\
+(\frac{E}{m}\tilde{D}_n^\parallel\hspace*{-2mm}&-&\hspace*{-2mm}\frac{\sqrt
{E^2-m^2}}{m}\tilde{B}_n^\parallel-\frac{E^2}{lm^2}\sum_p
B_{n-p}^\parallel B_p^\parallel)e^{-in(\sigma-\tau)}]+
{\cal O}(e^{-2R/l}).\nonumber
\end{eqnarray}
{}From the expressions of $(\eta^x,\;\eta^t,\;\eta^R)$ and
$(\xi^x,\;\xi^t,\;\xi^R),$
it follows that the center of mass solution and the first order perturbations
already give the complete solution in the asymptotic region. This is what it
should be since the black string background is asymptotically flat. We
therefore choose the initial conditions such that $\xi^x=\xi^t=\xi^R=0,$ i.e.
we take:
\begin{eqnarray}
&\tilde{A}_n^\perp=\tilde{B}_n^\perp=0,\;\;\;\;E\tilde{A}_n^\parallel=
\sqrt{E^2-m^2}\;\tilde{C}_n^\parallel,\;\;\;\;E\tilde{B}_n^\parallel=
\sqrt{E^2-m^2}\;\tilde{D}_n^\parallel,&\nonumber
\end{eqnarray}
\begin{eqnarray}
&\tilde{C}_n^\parallel=\frac{E^3}{lm^3}\sum_p
A_{n-p}^\parallel A_p^\parallel,\;\;\;\;\tilde{D}_n^\parallel=
\frac{E^3}{lm^3}\sum_p B_{n-p}^\parallel B_p^\parallel.&
\end{eqnarray}
Let us finally consider the world-sheet energy-momentum tensor that was
introduced in Eq.(2.7). Up to second order
in the expansion around the string center of mass we find:
\begin{eqnarray}
T_{\pm\pm}\hspace*{-2mm}&=&\hspace*{-2mm}-\frac{1}{4}m^2\alpha'^2+g_{\mu\nu}
\dot{q}^\mu\partial_\pm\eta^\nu
+\frac{1}{4}g_{\mu\nu,\rho}\dot{q}^\mu
\dot{q}^\nu\eta^\rho\nonumber\\
\hspace*{-2mm}&+&\hspace*{-2mm}g_{\mu\nu}\dot{q}^\mu\partial_\pm\xi^\nu
+g_{\mu\nu}\partial_\pm\eta^\mu\partial_\pm
\eta^\nu+g_{\mu\nu,\rho}\dot{q}^\mu\eta^\rho\partial_\pm\eta^\nu\nonumber\\
\hspace*{-2mm}&+&\hspace*{-2mm}\frac{1}{4}g_{\mu\nu,\rho}\dot{q}^\mu
\dot{q}^\nu\xi^\rho+\frac{1}{8}g_{\mu\nu,\rho\sigma}\dot{q}^\mu
\dot{q}^\nu\eta^\rho\eta^\sigma
\end{eqnarray}
Using the expressions for the first and second order perturbations,
Eqs.(4.35)-(4.36) and Eqs.(4.40)-(4.41), as
well as the conditions (4.45), it is straightforward now to
calculate $T_{++}$ and $T_{--}$ in the asymptotic region
$(r\rightarrow\infty)$:
\begin{equation}
T_{++}=-\frac{1}{4}m^2\alpha'^2-\sum_{R=\perp,\parallel}\;\sum_{n,p}
p(n-p)A_p^R A_{n-p}^R e^{-in(\tau+\sigma)},
\end{equation}
\begin{equation}
T_{--}=-\frac{1}{4}m^2\alpha'^2-\sum_{R=\perp,\parallel}\;\sum_{n,p}
p(n-p)B_p^R B_{n-p}^R e^{-in(\sigma-\tau)},
\end{equation}
and we get the usual flat spacetime constraints. The mass formula in
particular takes the form:
\begin{equation}
m^2\alpha'^2=4\sum_n n^2 \sum_{R=\perp,\parallel}A_n^R A_{-n}^R=
4\sum_n n^2 \sum_{R=\perp,\parallel}B_n^R B_{-n}^R.
\end{equation}
This is completely different from the result obtained for the $2+1$ black
hole anti de Sitter spacetime (compare with Eq.(2.16)) which is dual to the
black string. It should be stressed, however, that the formula (4.49) was
obtained in the asymptotic region $r\rightarrow\infty$ of the black string
background, and therefore does
not include any contribution from
the bounded solutions found when $E^2<m^2.$ The expression
(2.16), on the other hand, is general for the $2+1$ black
hole anti de Sitter spacetime.

For the black string, in the asymptotic region, we therefore obtain the density
of levels:
\begin{equation}
\rho(m)\sim e^{m\sqrt{\alpha'}},
\end{equation}
similar to the expression in flat Minkowski space, up to multiplication by
a polynomium in $m.$ The partition function for a gas of strings at
temperature $\beta^{-1},$ in the
asymptotic region of the black string background, therefore goes like:
\begin{equation}
Z(\beta)\sim\int^{\infty} dm\; e^{-m(\beta-\sqrt{\alpha'}\;)},
\end{equation}
which is only defined for $\beta>\sqrt{\alpha'},$ i.e. there is a Hagedorn
temperature:
\begin{equation}
T_{\mbox{Hg}}=(\alpha')^{-1/2}.
\end{equation}
In higher dimensional ($D\geq 4$) black hole spacetimes the next step now
would be to set up a scattering formalism, where a string from an asymptotic
in-state interacts with the gravitational field of the black hole and reappears
in an asymptotic out-state \cite{san7}. However, this is not possible in the
black string background. In the uncharged black string
background under consideration here, {\it all} null and timelike geodesics
incoming from spatial infinity pass through the horizon and fall into the
physical singularity \cite{hor}. No "angular momentum", as in the case of
scattering off the ordinary Schwarzschild black hole, can
prevent a point particle from falling into the singularity. The string
solutions considered in the present paper are based on perturbations around
the string center of mass which follows, at least approximately,
a point particle geodesic. A string incoming from spatial infinity therefore
inevitably falls into the singularity in the black string background.
\section{Concluding Remarks}
The classical string motion in anti de Sitter spacetime is stable in the sense
that it is oscillatory in time with real frequencies and the string
size and energy are bounded. Quantum mechanically, this reflects in the mass
operator, which is well defined for any value of the wave number $n,$ and
arbitrary high mass states (and therefore an infinite number of states) can
be constructed. This is to be contrasted with de Sitter spacetime, where
string instabilities develop, in the sense that the string size and energy
become unbounded for large de Sitter radius. For low mass states (the stable
regime), the mass operator in de Sitter spacetime is given by Eq.(1.1) but with
\begin{eqnarray}
\Omega_n(\lambda)_{\mbox{dS}}=
\frac{2n^2-m^2\alpha'\lambda}{\sqrt{n^2-m^2\alpha'\lambda}}.
\nonumber
\end{eqnarray}
Real mass solutions can be defined only up to some {\it maximal mass}
of the order $m^2\alpha'\approx 1/\lambda\;$ \cite{san1}.
For $\lambda<<1,$ real mass solutions can be defined
only for $N\leq N_{\mbox{max}}\sim 0.15/\lambda$ (where $N$ is the
eigenvalue of the number operator) and therefore there exists
a {\it finite} number of states only. These features of strings in
de Sitter spacetime have been recently confirmed within a different
(semi-classical) quantization
approach based on {\it exact} circular string solutions \cite{veg6}.

The presence of a cosmological constant $\Lambda$
(positive or negative) increases
considerably the number of levels of different eigenvalue of the
mass operator (there is a splitting of levels) with
respect to flat spacetime. That is, a non-zero cosmological
constant {\it decreases}
(although does not remove) the degeneracy of the string mass states,
introducing
a fine structure effect. For the low mass states the level spacing is
approximately constant (up to corrections of the order $\lambda^2$). For the
high mass states, the changes are more drastic and they depend crucially on
the sign of $\Lambda.$ A value $\Lambda<0$ causes the {\it growing}
of the level
spacing linearly with $N$ instead of being constant as in Minkowski
space. Consequently, the density of states $\rho(m)$ grows with the
exponential of $\sqrt{m}$ (instead of $m$ as in Minkowski space) discarding the
existence of a Hagedorn temperature in AdS spacetime, and the
possibility of a phase transition. In addition, another important feature of
the high mass string spectrum in AdS spacetime
is that it becomes independent of
$\alpha'.$ The string scale for the high mass states
is given by $\mid\Lambda\mid,$
instead of $1/\alpha'$ for the low mass states, as discussed at the end
of Section 3, Eqs.(3.32), (3.33).

The main physical features found in this paper are summarized in \\
Table 1.
\vskip 48pt
\hspace{-6mm}{\bf Acknowledgements:}\\
A.L.Larsen is supported by the Danish Natural Science Research Council under
Grant No. 11-1231-1SE.

\newpage
\begin{centerline}
{\bf Table Caption}
\end{centerline}
\vskip 24pt
\hspace*{-6mm}{\bf Table 1.} Characteristic
features of the quantum string mass spectrum
in anti de Sitter (AdS) and de Sitter (dS)
spacetimes. Notice the difference in the
high mass spectrum: In AdS the masses and level density become independent
of $\alpha'.$ In dS there is no such high mass spectrum at all.
\newpage
\begin{tabular}{|l|l|}\hline
\multicolumn{2}{|l|}{$ $}\\
\multicolumn{2}{|l|}{\hspace*{5.3cm} {\bf Quantum Strings} }\\
\multicolumn{2}{|l|}{$ $}\\
\hline\hline
$ $ & $ $ \\
\hspace*{7mm}Anti de Sitter spacetime (AdS) & \hspace*{1cm}de
Sitter spacetime (dS)\\
$ $ & $ $ \\ \hline
$ $ & $ $ \\
$ $ & Classical motion is unstable with\\
Classical motion is stable and & frequencies $\omega_n=\sqrt{n^2-
m^2\alpha'^2H^2}$\\
oscillatory in time with real & Unbounded string size and
energy\\
frequencies $\omega_n=\sqrt{n^2+m^2\alpha'^2H^2}$
& for large de Sitter radius, $R\rightarrow\infty.$\\
$ $ & $ $\\
$ $ & $ $\\
The mass formula is well defined & Real mass solutions only for \\
for all $m.$ There is an infinite & $m<1/(\alpha'H).$ Finite number of\\
number of states with arbitrary & states, $N_{\mbox{max}}\approx
0.15/(\alpha'H^2).$\\
high masses. $m^2\alpha'=0$ is an exact & $m^2\alpha'=0$ is an exact solution
at\\
solution at $D=25.$ & $D=25.$\\
$ $ & $ $\\
$ $ & $ $\\
The coupling to the gravitational & $ $\\
background produces a Fine structure & Fine structure effect appears in low\\
effect at all levels in the mass  & mass spectrum. Is similar to AdS; \\
spectrum. The number of levels & the differences appear to order\\
considerably increases with respect & $(\alpha'H^2)^3$\\
to flat space. & $ $\\
$ $ & $ $\\
For the high mass states: & $ $\\
$<m^2>\sim\mid\Lambda\mid N^2,\;\;\;J^2\sim m^2/\mid\Lambda\mid$ & $ $\\
Both are {\it independent} of $\alpha'\;$! & The similar region of high mass\\
The level spacing grows with $N.$ & states does not exist in the \\
$\rho(m)\sim \mbox{Exp}[m/\sqrt{\mid\Lambda\mid}\;)^{1/2}],\;\;$ high $m.$
& de Sitter spacetime.\\
No Hagedorn temperature exists. & $ $\\
$ $ & $ $\\ \hline
\end{tabular}
\vskip 12pt
\hspace*{63mm}{\bf Table 1}
\end{document}